# Magneto-chiral nonlinear optical effect with large anisotropic response in two-dimensional halide perovskite.

*Daichi Okada [1]\*,   Fumito Araoka[1]\**

[1] *RIKEN Center for Emergent Matter Science (CEMS), 2-1 Hirosawa, Wako, Saitama 351-0198, Japan*

ABSTRACT.

The chiral organic-inorganic halide perovskites (OIHPs) are vital candidates for superior nonlinear optical (NLO) effects associated with circularly polarized (CP) light. NLO in chiral materials often couples with magnetic dipole (MD) transition, as well as the conventional electric dipole (ED) transition. However, the importance of MD in NLO process of chiral OIHPs has not yet been well recognized. Here, the analysis of second harmonic generation circular dichroism (SHG-CD) provides the direct evidence that the MD contribution leads to a large anisotropic response to CP lights in chiral OIHPs, $(R-/S-MBACl)_2PbI_4$. The thin films exhibit great sensitivity to CP lights over a wide wavelength range, and the g-value reaches up to 1.57 at the wavelength where the contribution of MD is maximized. Furthermore, it is also effective as CP light generator, outputting CP-SHG with maximum g-factor of 1.76 upon the stimulation of linearly polarized light. This study deepens the understanding of the magneto-optical NLO processes in chiral systems.

Low dimensional organic-inorganic hybrid perovskites (OIHPs), in which perovskite and organic insulating layers are alternating, have higher exciton binding energy and better durability in air than three-dimensional (3D) OIHPs and are emerging as promising candidates for outstanding opto-electronic devices [1-4]. Low dimensional OIHPs are also attractive for its ability to incorporate various types of organic cations which tune the band gap, crystal structure, luminescence properties, and electrical characteristics. Such flexibility for structural design in low dimensional OIHP frameworks allows to incorporate chiral ligands into their crystalline lattice, generating novel concepts of materials based on the chiral OIHPs [5-7]. Meanwhile, chirality is a geometric description of the structure which cannot be superimposed on its mirror image. Chiral materials are able to recognize the handedness of circularly polarized (CP) lights or spin of electrons as manifested by an asymmetric behavior in optical absorption, emission, or photocurrent, that leads to unique optical activity, such as circular dichroism (CD) [8,9], CP luminescence [10-13] and detection [14-20], and spin filtering [21-26], those rarely seen in the conventional OIHPs. In addition, chirality introduces the broken symmetry, resulting in symmetry-dependent physical properties, such as even-order nonlinear optics [27-32], ferroelectricity [33-37], and (circular) photo-galvanic effect [38-42]. Thus, low-dimensional chiral OIHPs are promising materials that bring both the functional properties of perovskites and inherent chirality, since such materials must be essential in various innovative optoelectronic and spintronic devices.

CD, the dichroic response to the left- (L-) and right-handed (R-) CP lights, is general manifestation of optical activity in chiral substances. In most cases, the typical difference of absorption for L- and R-CP lights is less than 1%. Thus, even if one desires to utilize a chiral material to discriminate CP lights via CD, still it is far from use. On the other hand, optical activity also appears in nonlinear optical (NLO) responses in chiral materials, such as large asymmetric responses in optical second harmonics (SH) to CP fundamental lights. When chiral materials are stimulated with CP lights, the dichroic ratio in the SH signal is several orders of magnitude larger than the conventional CD in the linear regime. This optical activity in the NLO regime is the so-called second-harmonic generation circular dichroism (SHG-CD) and expected as a useful tool for efficient and sensitive detection of materials' chirality or handedness of CP lights [43-52]. The transition process of the linear optical activity is usually taken account by including the magnetic dipole (MD) transition [47,56]. Similarly, the NLO optical activity also arises from the contribution of the MD transition in addition to the electric dipole (ED) transition [43-48]. Recently, various types of chiral OIHPs have been explored to realize SHG-CD [49,50,53,54,55]. However, these reports have focused mainly on the efficiency of SHG or anisotropy of SHG-CD, and the detail of its transition process, in particular, contribution of the MD transition in SHG-CD, has not yet been well discussed.

In this study, we analyze SHG-CD in two dimensional (2D) chiral OIHPs and elucidate that the MD transition is the main contribution of the chiral NLO process in the present OIHPs. The 2D chiral OIHP thin films exhibit very large SHG-CD with a large dissymmetric factor (*g*-value) over a wide stimulation ranges from 1050 nm to 850 nm, reaching up to 1.57 as maximum. Quantitative analysis of SHG-CD using the quarter-wave plate (QWP) rotation method clearly proves the existence of the MD transition process and reveals the dominant contribution of MD leads to a larger asymmetric response of SHG-CD. Furthermore, we demonstrate the 2D chiral OHIPs as a coherent CP light generator via the NLO process. When the 2D chiral OHIP thin film is stimulated by a linearly polarized fundamental light, SHG is highly circularly polarized, and its *g*-value increases up to 1.76 at 455nm. Thus, the NLO response via the MD transition in 2D chiral OIHPs is useful for efficient discrimination and generation of CP lights.

The 2D chiral OIHPs containing R/S- 4-Chloro-α-methylbenzylamine (MBACl) were used in this study (Figure 1a). The crystals of (R/S-MBACl)$_2$PbI$_4$ are known to belong to the polar chiral group, P1, and exhibit ferroelectric and even-order NLO properties in addition to optical activity [33]. Single crystals of (R/S-MBACl)$_2$PbI$_4$, grown under cooling in a solution, were washed, and dissolved again in DMF, and then the solution was spin-casted and subsequently annealed to obtain thin films (~100nm) on quartz substrates. X-ray diffraction (XRD) data, in which only (0 0 2*l*) reflections were observed (Figure 1b), confirms the formation of 2D layers in polycrystalline domains, highly oriented parallel to the substrate. Note that XRD detected no remarkable difference between R and S, nor impurity peaks. Either the (R- or S-MBACl)$_2$PbI$_4$ thin film had a sharp excitonic absorption peak (Figure 1c) at about 495 nm in its ultraviolet-visible (UV-VIS) absorption spectrum. The CD spectra for (R/S- MBACl)$_2$PbI$_4$ films showed mirrored curves (Figure 1d) peaking at 495 nm corresponding to the above-mentioned excitonic resonance of the perovskite, which confirms that the chirality of the organic ligands influences the electronic property of the perovskite layers. The dissymmetric factor (*g*-value) for the CD peak at the excitonic resonance is $g_{CD} = 2.8 \times 10^{-4}$ (Figure S1). A racemic sample, (rac-MBACl)$_2$PbI$_4$, that is a 50:50 mixture of R and S enantiomers, is optically inactive (Figure 1d).

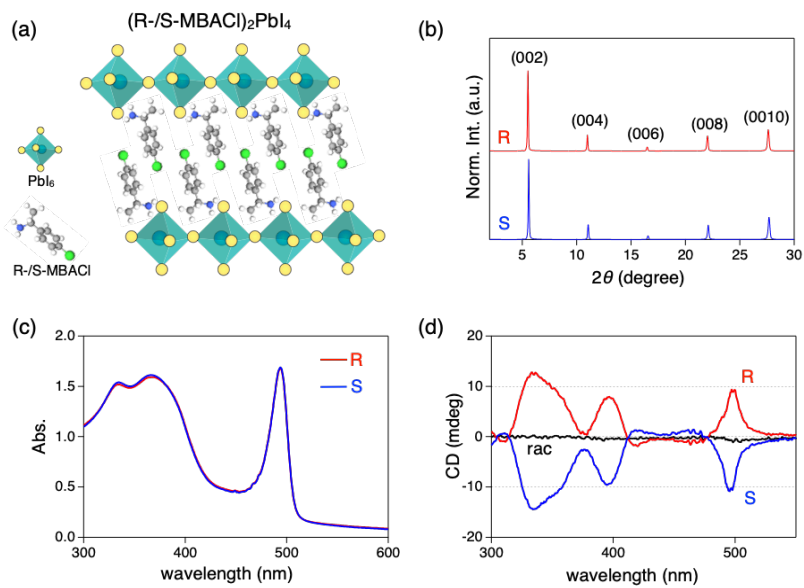

**Figure 1**, (a) Schematic representation of the two-dimensional chiral organic-inorganic hybrid perovskite, (R-/S-MBACl)$_2$PbI$_4$. (b) XRD profiles, (c) UV-VIS absorption and (d) CD spectra of (R-/S-MBACl)$_2$PbI$_4$ thin films. In (b-d), the data for the R- and S- sample films are shown with red and blue curves. In (d), the CD spectrum for the racemic sample (rac) is presented together (black line).

Upon the stimulation by focused femto-second laser pulses, SHG in the (R/S-MBACl)$_2$PbI$_4$ thin films was examined. The SHG signals were collected in the transmission configuration with a home-built optical setup using a tunable near-infrared laser (850 ~1050 nm, Chameleon, Coherent) (Figure 2a, detailed in Figure S2) and recorded in steps of 10 nm of fundamental stimulation wavelength. The laser was *p*-polarized and incident at 45° to the sample normal. Figure 2b shows a normalized SHG spectrum when stimulated at 120 mW. Both the R- and S- sample films show the similar dependences on the stimulation strength, that is, quadratic increase with the fundamental light intensity, which is a typical feature of SHG (Figure S3). Since the 2D OIHPs have a strong excitonic characteristic due to the spontaneous formation of multiple layered quantum well-like structure [1-4], the SHG efficiency is significantly increased near the excitonic absorption peak. However, the self-absorption of SHG is also increased and not negligible in this condition, and as a result the signal peak of SHG is slightly red-shifted from the absorption peak. The sample films were well optically isotropic, which is consistent with the fact that the SHG signal exhibited no significant dependence on the polarization angle of the fundamental light (Figure S4).

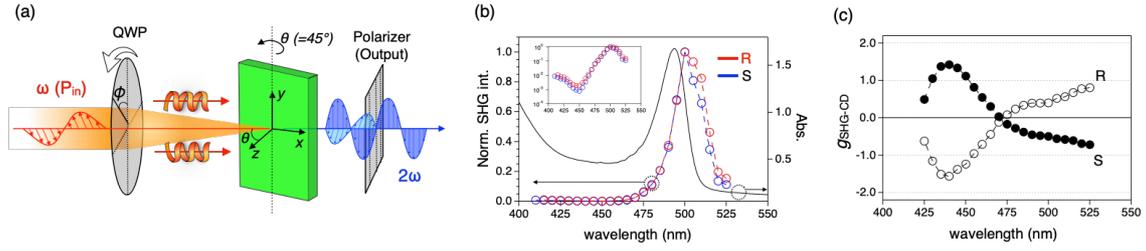

**Figure 2,** (a) Schematic of the experimental setup for SHG. (b) Normalized SHG intensity spectra of R- (red) and S-sample (blue) films, together with the UV-VIS absorption spectrum of the S-sample film. The inset in (b) shows the SHG spectra plotted on a log scale. (c)Spectra of the dissymmetric factors, $g_{SHG\text{-}CD}$, for SHG-CD without an output polarizer.

By setting the QWP at $\pm 45°$ with respect to the *p*-polarization direction, we can choose a L- or R-CP state for the fundamental light. Figure 2c plots the wavelength dependences of the SHG dissymmetry factor, $g_{SHG-CD} = 2\left(I_{\omega,L}^{2\omega,nonpol} - I_{\omega,R}^{2\omega,nonpol}\right)/\left(I_{\omega,L}^{2\omega,nonpol} + I_{\omega,R}^{2\omega,nonpol}\right)$, where $I_{\omega,L/R}^{2\omega,nonpol}$ refers to the non-polarized SHG intensity for the L- or R-CP fundamental light, respectively. As seen, the $g_{SHG-CD}$ spectra from the R- and S- sample films show large anisotropic response over a wide wavelength range and clear inversion between them. The maximum value of $|g_{SHG-CD}|$ is obtained at about 440 nm as 1.57/1.42 in R/S-sample films (Figure 2c), which is more than three orders of magnitude larger than the usual dissymmetry factor $g_{CD}$ in the linear CD (Figure S1) and one of the most efficient chiral NLO response in the OIHPs reported to date. The sign and sign inversion wavelength of $g_{SHG-CD}$ are reasonably associated with those of $g_{CD}$. However, interestingly, the $g_{SHG-CD}$ didn't show peaking at the excitonic resonance of the perovskite (495 nm) and increased towards lower wavelengths in the off-resonance region, which perhaps means scarce correlation with the linear optical regime. On the other hand, the racemic sample film didn't show SHG-CD at all (Figure S5).

In order to clarify the origin of such a large chiral NLO response, we performed the so-called QWP rotation analysis, by which the asymmetric response in SHG was quantitatively analyzed by continuous modulation of the polarization of the fundamental light upon rotating the QWP. Note that in this analysis, optical isotropy should be guaranteed, because anisotropy may add geometrical chirality which causes an essential problem in such polarization analysis [57,58]. This in turn means, the afore-mentioned optical isotropy in our present samples is ideal for pursuit of the pure chiral optical effect using this method. For this analysis, an output polarizer was placed behind the sample to resolve the light field of SHG into the *p*- and *s*-polarized components (Figure 2a). In general, the SHG intensity $I(2\omega)$ can be expressed in terms of the *p*- and *s*-polarized complex components, $E_p(\omega)$ and $E_s(\omega)$, of the fundamental light [43-48],

$$I(2\omega) = \left[l(\theta)E_p(\omega)E_p(\omega) + m(\theta)E_s(\omega)E_s(\omega) + n(\theta)E_p(\omega)E_s(\omega)\right]^2 \quad (1)$$

where $l(\theta)$, $m(\theta)$, and $n(\theta)$ are complex factors, each of which is a linear combination of the 2nd-order nonlinear susceptibility tensor components as a function of the incident angle, $\theta$ (= 45° in the present study). Since $E_p(\omega)$ and $E_s(\omega)$ are functions of the rotation angle, $\varphi$, of the QWP, Eq. (1) is rewritten in terms of the complex fields, $L(\varphi)$, $M(\varphi)$ and $N(\varphi)$, as follows,[43-48].

$$I(2\omega) = [l(\theta)L(\varphi) + m(\theta)M(\varphi) + n(\theta)N(\varphi)]^2 \quad (2)$$
$$L(\varphi) = E_0^2(\sin^2\varphi + i\cos^2\varphi)^2$$
$$M(\varphi) = E_0^2\sin^2\varphi\cos^2\varphi(1-i)^2$$
$$N(\varphi) = E_0^2(\sin^2\varphi + i\cos^2\varphi)\sin\varphi\cos\varphi(1-i)$$

Here, the chiral nonlinear system needs to be considered with, in addition to the conventional ED-only induced nonlinear polarization process, two more transition processes, that are the MD induced nonlinear polarization and the ED induced nonlinear magnetization. Thus, the nonlinear susceptibility tensors are introduced as $\chi^{eee}$, $\chi^{eem}$ and $\chi^{mee}$, which are corresponding to the above three processes, respectively (detailed in Figure S6). In the case of optically isotropic chiral systems (in particular, of $C_\infty$ symmetry in the present case), the nonlinear susceptibility tensor components are eliminated. Then, $l(\theta)$, $m(\theta)$, and $n(\theta)$ are expanded using these tensor elements depending on the output polarizations [43-48],

$$l_p = \sin\theta\,[\chi^{eee}_{zzz}\sin^2\theta + \chi^{eee}_{zxx}\cos^2\theta + 2\chi^{eee}_{xxz}\cos^2\theta - \chi^{eem}_{zxy}\cos\theta - \chi^{eem}_{zxy}\cos\theta$$
$$+ 2\chi^{mee}_{xyz}\cos\theta]$$

$$m_p = \sin\theta\,(\chi^{eee}_{zxx} - \chi^{eem}_{zxy}\cos\theta + \chi^{eem}_{xyz}\cos\theta)$$

$$n_p = \sin\theta\,[2\chi^{eee}_{xyz}\cos\theta + \left(\chi^{eem}_{zzz} - \chi^{eem}_{zxx}\right)\sin^2\theta + \left(\chi^{eem}_{xzx} - \chi^{eem}_{xxz}\right)\cos^2\theta - 2\chi^{mee}_{xxz}]$$

$$l_s = \sin\theta\,[-2\chi^{eee}_{xyz}\cos\theta - \chi^{eem}_{xzx} + \chi^{mee}_{zzz}\sin^2\theta + \chi^{mee}_{zxx}\cos^2\theta + 2\chi^{mee}_{xxz}\cos^2\theta]$$

$$m_s = \sin\theta\,(\chi^{eem}_{xxz} + \chi^{mee}_{zxx})$$

$$n_s = \sin\theta\,[2\chi^{eee}_{xxz} - \left(\chi^{eem}_{xzy} + \chi^{eem}_{xyz}\right)\cos\theta + 2\chi^{mee}_{xyz}\cos\theta] \quad (3)$$

where the subscripts $p$ and $s$ represent the polarization of the SHG light (Figure 2a). According to Eq. (3), SHG should be deactivated in case the incident laser angle is 0°, that is experimentally confirmed (Figure S7). As seen in Eq. (2), the circular difference in SHG is originated from in expanded term $N(\varphi)$, since only this function has a different sign for the two circular polarizations (i.e. rotation angles of $\varphi = \pm 45°$). Thus, the imaginary parts of $n_p$ and $n_s$ are the main contribution in SHG-CD and the magnitude of anisotropic response is linearly related to $\text{Im}[(l-m)n^*]$.[43-48] Furthermore, the $n_p$ and $l_s$, $m_s$ are only allowed to take finite value when the material possess chirality. In principle, SHG-CD can occur even only by the ED transition process due to the $\chi^{eee}$ terms in $n_p$ and $n_s$. However, the imaginary part of $\chi^{eee}$ is usually negligibly small in off-resonant conditions, so that SHG-CD upon the ED approximation is hardly observed in the wavelength regions far from the absorption band. On the other hand, in the case of the MD transition process, it happens oppositely, i.e. $\chi^{mee}$ and $\chi^{eem}$ take imaginary values in the optically off-resonant condition and thus pronounced [43-48]. For this reason, it is natural to consider that the observed large SHG-CD at the wavelength regions detuned from the excitonic resonance is attributed to the MD transition process in the NLO regime (Figure 2c). Additionally, it is important to note that $m_s$ depends only on $\chi^{eem}_{xxz}$ and $\chi^{mee}_{zxx}$ of the MD transition process and hence must be null under the ED approximation irrespectively of being resonant or not. Therefore, non-zero $m_s$ must be direct evidence of the MD transition process [46]. Figure 3a shows *p*- and *s*-polarized SHG signals in R- and S-sample films at 440 nm, plotted as a function of $\varphi$ together with the best fitting curves using Eq. (2). The estimated relative values of $l_{p/s}$, $m_{p/s}$, and $n_{p/s}$ at 440 nm, normalized by achiral index are compared in Table 1. Note that $n_p$ and $l_s$, $m_s$ take finite values significantly out of phase with achiral index. Their signs are opposite but almost the same in magnitude between the R- and S-sample films. As already mentioned, the finite $m_s$ clearly proves the MD transition process in the NLO response. We also confirmed the similar tendency for other wavelength regions (Figure S8, Table S1), i.e. the clear SHG-CD performance is observed in the entire stimulation range, even though SHG polarization is resolved to *p*- and *s*-polarization (Figure S9). Here, we collected the data at incident laser angle of 45˚, but $g_{\text{SHG-CD}}$ value is independent to incident laser angle except at 0˚, indicating that MD transition is also not influenced by it (Figure S10).

To discuss the relationship between the dissymmetric factor $g_{SHG-CD}$ and the MD transition, we examined the wavelength dependence of $m_s$ in the S-sample film. Figure 3c shows the plot of the real and imaginary components of $m_s$. Figure 3d and 3e are $|m_s|$ and $|g_{SHG-CD}|$, respectively, where the spectral shapes show similar tendency, i.e. the larger MD contribution gives rise to larger $|g_{SHG-CD}|$ values, proving the MD transition critically leads to the highly efficient discrimination of CP lights. $|n_p|$ also has the similar trend with $|m_s|$ (Figure S11). In contrast, the values of $l_s$, $m_s$, and $n_p$ in the racemic sample film are very small and almost zero at any observed wavelengths, which also confirms that the finite $l_s$, $m_s$, and $n_p$ in the chiral OHIP samples are resulted from the material's chirality (Figure S8, S11, Table1, S1). It is noteworthy that $m_s$ is insignificant at the resonance condition, while it is rather pronounced in the off-resonant conditions. This is plausibly because $\chi_{xxz}^{eee}$ from the ED transition is reduced, and the MD contribution becomes relatively more significant. Accordingly, in off-resonant conditions, the SHG intensity largely decreases due to the diminished ED contribution, while the MD contribution is stressed to yield a large chiral NLO response. For example, in the case of the R-sample film, although the largest $g_{SHG-CD}$ of 1.57 is obtained at the SHG wavelength of 440 nm, the SHG intensity is decreased by three orders of magnitude compared to the maximum (Figure 2b). The systematic investigation of wavelength dependence of SHG-CD revealed a clear correlation between MD and $g_{SHG-CD}$, allowing us to identify the wavelength that gives an exceptionally high $g_{SHG-CD}$. By predomination of the MD contribution in the NLO process, we have successfully achieved significantly higher $g_{SHG-CD}$ than those in the previously reported chiral OIHPs. Until now, the evaluation of SHG-CD in chiral OIHPs has been limited to selected wavelengths, and hence the potential of dissymmetric chiral NLO responses has not been sufficiently discussed nor examined. Thus, our present research also suggests an importance of comprehensive study on the dissymmetric chiral NLO response.

**Table 1,** Relative fitting value of $l$, $m$, $n$ for R-, S-and (rac-MBACl)$_2$PbI$_4$ at $\lambda_{SHG}$= 440nm

|     | $l_p$ | $m_p$        | $n_p$ (chiral)  | $l_s$ (chiral)  | $m_s$ (chiral)  | $n_s$ |
| --- | ----- | ------------ | --------------- | --------------- | --------------- | ----- |
| R   | 1     | -0.04-i0.60  | -(1.32-i0.31)   | -(0.80+i1.22)   | 0.48+i0.51      | 1     |
| S   | 1     | -0.26-i0.72  | 1.31-i0.28      | 0.78+i1.14      | -(0.32+i0.54)   | 1     |
| rac | 1     | -0.43-i0.40  | 0               | 0.03            | -0.04-i0.01     | 1     |

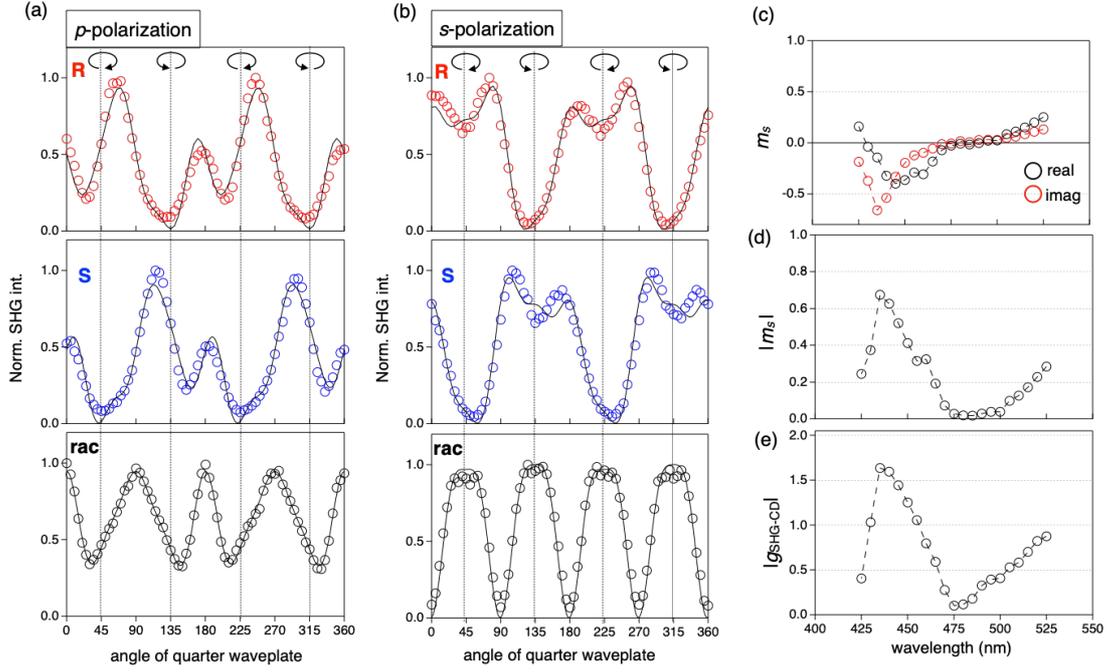

**Figure 3,** (a, b) Normalized SHG intensity plot as function of rotation angle of QWP in *p*-(a) and *s*- (b) polarized conditions for R- (Top), S-(Middle) and (rac-MBACl)$_2$PbI$_4$ (Bottom) at $\lambda_{SHG}$= 440nm. The black solid line is fitting curve by using Eq. (2). (c, d) The spectrum of $m_s$ with real and imaginary component (c) and absolute value of $m_s$ ($|m_s|$) (d). (e) The spectrum of $|g_{SHG-CD}|$ in *s*-polarized condition.

We also evaluated circularly polarized second harmonic generation (CP-SHG) in these R- and S-sample films to demonstrate that the chiral NLO is also useful for coherent CP light generation *via* the NLO process. In the case of CP-SHG, we use the *p*-polarized fundamental light, and the degree of circular polarization of the SHG light field is analyzed using a QWP and a polarizer behind the sample (Figure 4a). It was revealed that the output SHG is highly circularly polarized and shows inverted responses in R- and S-sample films (Figure 4b). It is worth recalling that the sample is optically isotropic. Figure 4c shows a wavelength dependence of the dissymmetric factor for CP-SHG, $g_{CP-SHG} = 2\left(I^{2\omega,L}_{\omega,p} - I^{2\omega,R}_{\omega,p}\right)/\left(I^{2\omega,L}_{\omega,p} + I^{2\omega,R}_{\omega,p}\right)$, where $I^{2\omega,L/R}_{\omega,p}$ is defined as an intensity of L- or R-CP-SHG signal, differently from $g_{SHG-CD}$. The R- and S-sample films gave $g_{CP-SHG}$ = -1.55 and 1.76 at 455 nm, which are several orders of magnitude larger than the conventional CP luminescence from chiral fluorescent molecular systems [59-61]. Moreover, since SHG is a coherent wavelength conversion process, the output CP-SHG light is also coherent, suggesting the possibility for various optical applications, such as light sources or converters for optical communication or holographic laser displays, etc. Thus, the NLO process in the chiral OHIP is a powerful tool to efficiently generate coherent CP lights.

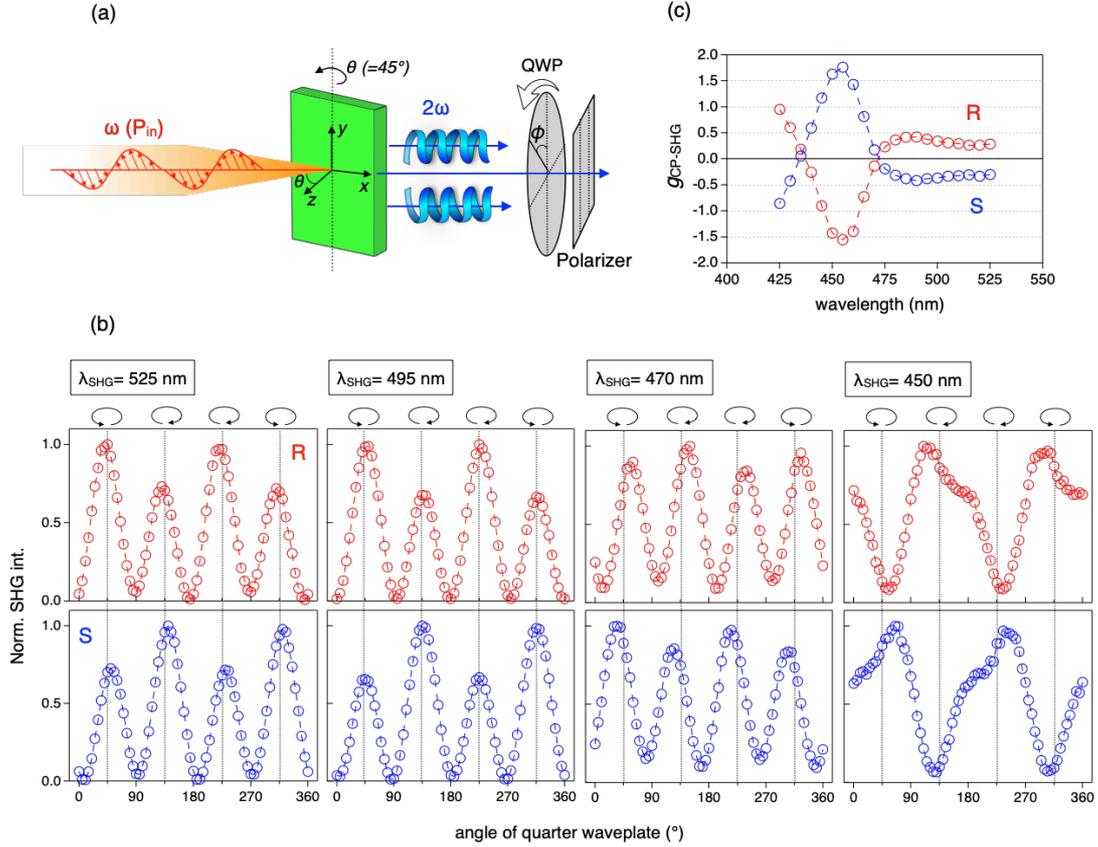

**Figure 4**, (a) Schematic representation of setup for detection of CP-SHG signal. The polarizer is placed on the detection side with cross polarizing to incident laser polarization. (b) The CP-SHG plot as function of angle of QWP of R (red) and S (blue) at several SHG wavelength. (c) The $g_{CP\text{-}SHG}$ spectrum of R (red) and S (blue)

In summary, we reveal the importance of MD contribution to the chiral NLO response in the two-dimensional OIHPs, (R-/S-MBACl)$_2$PbI$_4$ by utilizing systematic SHG-CD analysis. MD transition promotes a phase difference between chiral and achiral tensor components, resulting in a large anisotropic response to CP lights that can reach a maximum $g_{SHG\text{-}CD}$ of 1.57 in (R-MBACl)$_2$PbI$_4$ thin film. Moreover, (R-/S-MBACl)$_2$PbI$_4$ demonstrates excellent CP lights generator, with a $g_{CP\text{-}SHG}$ of up to 1.76 in the CP-SHG process. The discrimination and generation of CP lights with such high g-factors represent the highest-level performance observed among chiral materials so far. However, the present study has not achieved both high efficiency SHG and large $g_{SHG\text{-}CD}$, which remains a challenge for future work. This study not only advances our understanding of the NLO and magneto-optics in chiral OIHPs but also demonstrates that chiral OIHPs are advantageous for optical quantum information devices and highly throughput chiral sensing.

AUTHOR INFORMATION

**Corresponding Author**

**Daichi Okada**: RIKEN Center for Emergent Matter Science (CEMS),

2-1 Hirosawa, Wako, Saitama 351-0198, Japan

E-mail: daichi.okada @riken.jp

**Fumito Araoka**: RIKEN Center for Emergent Matter Science (CEMS),

2-1 Hirosawa, Wako, Saitama 351-0198, Japan

E-mail: fumito.araoka@riken.jp

**Author Contributions**

D. O and F. A contributed equally this work.

**Funding Sources**

JSPS KAKENHI (JP 23H01942, JP 21K14605, JP21H01801), JST CREST (JPMJCR17N1, JPMJCR23O1), JST SICORP EIG CONCERT-Japan (JPMJSC22C3)

**Notes**

The authors declare no competing financial interest.

ACKNOWLEDGMENT

This work was partly supported by Grant-in-Aid for Scientific research (B) (JP 23H01942, JP21H01801) and Young Scientists (JP21K14605) from Japan Society for the Promotion of Science (JSPS), JST CREST (JPMJCR17N1, JPMJCR23O1), and JST SICORP EIG CONCERT-Japan (JPMJSC22C3).

REFERENCES

(1) Chen, Y.; Sun, Y.; Peng, J.; Tang, J.; Zheng, K.; Liang, Z. 2D Ruddlesden-Popper Perovskites for Optoelectronics. *Adv. Mater.* **2018**, *30* (2), 1703487.

(2) Hu, J.; Yan, L.; You, W. Two-Dimensional Organic–Inorganic Hybrid Perovskites: A New Platform for Optoelectronic Applications. *Adv. Mater.* **2018**, *30* (48), 1802041.

(3) Katan, C.; Mercier, N.; Even, J. Quantum and Dielectric Confinement Effects in Lower-Dimensional Hybrid Perovskite Semiconductors. *Chem. Rev.* **2019**, *119* (5), 3140–3192.

(4) Blancon, J.-C.; Even, J.; Stoumpos, Costas. C.; Kanatzidis, Mercouri. G.; Mohite, A. D. Semiconductor Physics of Organic–Inorganic 2D Halide Perovskites. *Nat. Nanotechnol.* **2020**, *15*

(12), 969–985.

(5) Long, G.; Sabatini, R.; Saidaminov, M. I.; Lakhwani, G.; Rasmita, A.; Liu, X.; Sargent, E. H.; Gao, W. Chiral-Perovskite Optoelectronics. *Nat Rev Mater* **2020**, *5* (6), 423–439.

(6) Ma, S.; Ahn, J.; Moon, J. Chiral Perovskites for Next-Generation Photonics: From Chirality Transfer to Chiroptical Activity. *Adv. Mater.* **2021**, *33* (47), 2005760.

(7) Ma, J.; Wang, H.; Li, D. Recent Progress of Chiral Perovskites: Materials, Synthesis, and Properties. *Adv. Mater.* **2021**, *33* (26), 2008785.

(8) Ahn, J.; Lee, E.; Tan, J.; Yang, W.; Kim, B.; Moon, J. A New Class of Chiral Semiconductors: Chiral-Organic-Molecule-Incorporating Organic–Inorganic Hybrid Perovskites. *Mater. Horiz.* **2017**, *4* (5), 851–856.

(9) Ahn, J.; Ma, S.; Kim, J.-Y.; Kyhm, J.; Yang, W.; Lim, J. A.; Kotov, N. A.; Moon, J. Chiral 2D Organic Inorganic Hybrid Perovskite with Circular Dichroism Tunable Over Wide Wavelength Range. *J. Am. Chem. Soc.* **2020**, *142* (9), 4206–4212.

(10) Long, G.; Jiang, C.; Sabatini, R.; Yang, Z.; Wei, M.; Quan, L. N.; Liang, Q.; Rasmita, A.; Askerka, M.; Walters, G.; Gong, X.; Xing, J.; Wen, X.; Quintero-Bermudez, R.; Yuan, H.; Xing, G.; Wang, X. R.; Song, D.; Voznyy, O.; Zhang, M.; Hoogland, S.; Gao, W.; Xiong, Q.; Sargent, E. H. Spin Control in Reduced-Dimensional Chiral Perovskites. *Nature Photon* **2018**, *12* (9), 528–533.

(11) Ma, J.; Fang, C.; Chen, C.; Jin, L.; Wang, J.; Wang, S.; Tang, J.; Li, D. Chiral 2D Perovskites with a High Degree of Circularly Polarized Photoluminescence. *ACS Nano* **2019**, *13* (3), 3659–3665.

(12) Di Nuzzo, D.; Cui, L.; Greenfield, J. L.; Zhao, B.; Friend, R. H.; Meskers, S. C. J. Circularly Polarized Photoluminescence from Chiral Perovskite Thin Films at Room Temperature. *ACS Nano* **2020**, *14* (6), 7610–7616.

(13) Zhan, G.; Zhang, J.; Zhang, L.; Ou, Z.; Yang, H.; Qian, Y.; Zhang, X.; Xing, Z.; Zhang, L.; Li, C.; Zhong, J.; Yuan, J.; Cao, Y.; Zhou, D.; Chen, X.; Ma, H.; Song, X.; Zha, C.; Huang, X.; Wang, J.; Wang, T.; Huang, W.; Wang, L. Stimulating and Manipulating Robust Circularly Polarized Photoluminescence in Achiral Hybrid Perovskites. *Nano Lett.* **2022**, *22* (10), 3961–3968.

(14) Chen, C.; Gao, L.; Gao, W.; Ge, C.; Du, X.; Li, Z.; Yang, Y.; Niu, G.; Tang, J. Circularly Polarized Light Detection Using Chiral Hybrid Perovskite. *Nat Commun* **2019**, *10* (1), 1927.

(15) Wang, L.; Xue, Y.; Cui, M.; Huang, Y.; Xu, H.; Qin, C.; Yang, J.; Dai, H.; Yuan, M. A Chiral Reduced-Dimension Perovskite for an Efficient Flexible Circularly Polarized Light Photodetector. *Angew. Chem. Int. Ed.* **2020**, *59* (16), 6442–6450.

(16) Ishii, A.; Miyasaka, T. Direct Detection of Circular Polarized Light in Helical 1D Perovskite-Based Photodiode. *Sci. Adv.* **2020**, *6* (46), eabd3274.

(17) Peng, Y.; Liu, X.; Li, L.; Yao, Y.; Ye, H.; Shang, X.; Chen, X.; Luo, J. Realization of Vis–NIR Dual-Modal Circularly Polarized Light Detection in Chiral Perovskite Bulk Crystals. *J. Am. Chem. Soc.* **2021**, *143* (35), 14077–14082.

(18) Zhang, X.; Liu, X.; Li, L.; Ji, C.; Yao, Y.; Luo, J. Great Amplification of Circular Polarization Sensitivity via Heterostructure Engineering of a Chiral Two-Dimensional Hybrid Perovskite Crystal with a Three-Dimensional MAPbI$_3$ Crystal. *ACS Cent. Sci.* **2021**, *7* (7), 1261–1268.

(19) Liu, T.; Shi, W.; Tang, W.; Liu, Z.; Schroeder, B. C.; Fenwick, O.; Fuchter, M. J. High Responsivity Circular Polarized Light Detectors Based on Quasi Two-Dimensional Chiral Perovskite Films. *ACS Nano* **2022**, *16* (2), 2682–2689.

(20) Yao, B.; Wei, Q.; Yang, Y.; Zhou, W.; Jiang, X.; Wang, H.; Ma, M.; Yu, D.; Yang, Y.; Ning, Z. Symmetry-Broken 2D Lead–Tin Mixed Chiral Perovskite for High Asymmetry Factor Circularly Polarized Light Detection. *Nano Lett.* **2023**, *23* (5), 1938–1945.

(21) Lu, H.; Wang, J.; Xiao, C.; Pan, X.; Chen, X.; Brunecky, R.; Berry, J. J.; Zhu, K.; Beard, M. C.; Vardeny, Z. V. Spin-Dependent Charge Transport through 2D Chiral Hybrid Lead-Iodide Perovskites. *Sci. Adv.* **2019**, *5* (12), eaay0571.

(22) Lu, H.; Xiao, C.; Song, R.; Li, T.; Maughan, A. E.; Levin, A.; Brunecky, R.; Berry, J. J.; Mitzi, D. B.; Blum, V.; Beard, M. C. Highly Distorted Chiral Two-Dimensional Tin Iodide Perovskites for Spin Polarized Charge Transport. *J. Am. Chem. Soc.* **2020**, *142* (30), 13030–13040.

(23) Kim, Y.-H.; Zhai, Y.; Lu, H.; Pan, X.; Xiao, C.; Gaulding, E. A.; Harvey, S. P.; Berry, J. J.; Vardeny, Z. V.; Luther, J. M.; Beard, M. C. Chiral-Induced Spin Selectivity Enables a Room-Temperature Spin Light-Emitting Diode. *Science* **2021**, *371* (6534), 1129–1133.

(24) Wang, J.; Lu, H.; Pan, X.; Xu, J.; Liu, H.; Liu, X.; Khanal, D. R.; Toney, M. F.; Beard, M. C.; Vardeny, Z. V. Spin-Dependent Photovoltaic and Photogalvanic Responses of Optoelectronic Devices Based on Chiral Two-Dimensional Hybrid Organic–Inorganic Perovskites. *ACS Nano* **2021**, *15* (1), 588–595.

(25) Lu, Y.; Wang, Q.; Chen, R.; Qiao, L.; Zhou, F.; Yang, X.; Wang, D.; Cao, H.; He, W.; Pan, F.; Yang, Z.; Song, C. Spin-Dependent Charge Transport in 1D Chiral Hybrid Lead-Bromide Perovskite with High Stability. *Adv. Funct. Mater.* **2021**, *31* (43), 2104605.

(26) Maiti, A.; Pal, A. J. Spin-Selective Charge Transport in Lead-Free Chiral Perovskites: The Key towards High-Anisotropy in Circularly-Polarized Light Detection. *Angew Chem Int Ed* **2022**, *61* (52), e20221416.

(27) Dehnhardt, N.; Axt, M.; Zimmermann, J.; Yang, M.; Mette, G.; Heine, J. Band Gap-Tunable, Chiral Hybrid Metal Halides Displaying Second-Harmonic Generation. *Chem. Mater.* **2020**, *32* (11), 4801–4807.

(28) Fu, D.; Xin, J.; He, Y.; Wu, S.; Zhang, X.; Zhang, X.; Luo, J. Chirality-Dependent Second-Order Nonlinear Optical Effect in 1D Organic–Inorganic Hybrid Perovskite Bulk Single Crystal.

*Angew Chem Int Ed* **2021**, *60* (36), 20021–20026.

(29) Zhao, L.; Han, X.; Zheng, Y.; Yu, M.-H.; Xu, J. Tin-Based Chiral Perovskites with Second-Order Nonlinear Optical Properties. *Adv Photo Res* **2021**, *2* (11), 2100056.

(30) Zhao, J.; Zhao, Y.; Guo, Y.; Zhan, X.; Feng, J.; Geng, Y.; Yuan, M.; Fan, X.; Gao, H.; Jiang, L.; Yan, Y.; Wu, Y. Layered Metal-Halide Perovskite Single-Crystalline Microwire Arrays for Anisotropic Nonlinear Optics. *Adv Funct Materials* **2021**, *31* (48), 2105855.

(31) Yao, L.; Zeng, Z.; Cai, C.; Xu, P.; Gu, H.; Gao, L.; Han, J.; Zhang, X.; Wang, X.; Wang, X.; Pan, A.; Wang, J.; Liang, W.; Liu, S.; Chen, C.; Tang, J. Strong Second- and Third-Harmonic Generation in 1D Chiral Hybrid Bismuth Halides. *J. Am. Chem. Soc.* **2021**, *143* (39), 16095–16104.

(32) Ge, F.; Li, B.; Cheng, P.; Li, G.; Ren, Z.; Xu, J.; Bu, X. Chiral Hybrid Copper(I) Halides for High Efficiency Second Harmonic Generation with a Broadband Transparency Window. *Angew Chem Int Ed* **2022**, *61* (10), e2021150.

(33) Yang, C.; Chen, W.; Ding, Y.; Wang, J.; Rao, Y.; Liao, W.; Tang, Y.; Li, P.; Wang, Z.; Xiong, R. The First 2D Homochiral Lead Iodide Perovskite Ferroelectrics: [ *R* - and *S* -1-(4-Chlorophenyl) Ethylammonium] $_2$ PbI $_4$. *Adv. Mater.* **2019**, *31* (16), 1808088.

(34) Gao, J.-X.; Zhang, W.-Y.; Wu, Z.-G.; Zheng, Y.-X.; Fu, D.-W. Enantiomorphic Perovskite Ferroelectrics with Circularly Polarized Luminescence. *J. Am. Chem. Soc.* **2020**, *142* (10), 4756–4761.

(35) Hu, Y.; Florio, F.; Chen, Z.; Phelan, W. A.; Siegler, M. A.; Zhou, Z.; Guo, Y.; Hawks, R.; Jiang, J.; Feng, J.; Zhang, L.; Wang, B.; Wang, Y.; Gall, D.; Palermo, E. F.; Lu, Z.; Sun, X.; Lu, T.-M.; Zhou, H.; Ren, Y.; Wertz, E.; Sundararaman, R.; Shi, J. A Chiral Switchable Photovoltaic Ferroelectric 1D Perovskite. *Sci. Adv.* **2020**, *6* (9), eaay4213.

(36) Li, L.-S.; Tan, Y.-H.; Wei, W.-J.; Gao, H.-Q.; Tang, Y.-Z.; Han, X.-B. Chiral Switchable Low-Dimensional Perovskite Ferroelectrics. *ACS Appl. Mater. Interfaces* **2021**, *13* (1), 2044–2051.

(37) Zeng, Y.; Huang, X.; Huang, C.; Zhang, H.; Wang, F.; Wang, Z. Unprecedented 2D Homochiral Hybrid Lead-Iodide Perovskite Thermochromic Ferroelectrics with Ferroelastic Switching. *Angew. Chem. Int. Ed.* **2021**, *60* (19), 10730–10735.

(38) Huang, P.-J.; Taniguchi, K.; Miyasaka, H. Bulk Photovoltaic Effect in a Pair of Chiral–Polar Layered Perovskite-Type Lead Iodides Altered by Chirality of Organic Cations. *J. Am. Chem. Soc.* **2019**, *141* (37), 14520–14523.

(39) Wang, J.; Lu, H.; Pan, X.; Xu, J.; Liu, H.; Liu, X.; Khanal, D. R.; Toney, M. F.; Beard, M. C.; Vardeny, Z. V. Spin-Dependent Photovoltaic and Photogalvanic Responses of Optoelectronic Devices Based on Chiral Two-Dimensional Hybrid Organic–Inorganic Perovskites. *ACS Nano* **2021**, *15* (1), 588–595.

(40) Huang, P.; Taniguchi, K.; Shigefuji, M.; Kobayashi, T.; Matsubara, M.; Sasagawa, T.; Sato,

H.; Miyasaka, H. Chirality-Dependent Circular Photogalvanic Effect in Enantiomorphic 2D Organic–Inorganic Hybrid Perovskites. *Adv. Mater.* **2021**, *33* (17), 2008611.

(41) Fan, C.; Han, X.; Liang, B.; Shi, C.; Miao, L.; Chai, C.; Liu, C.; Ye, Q.; Zhang, W. Chiral Rashba Ferroelectrics for Circularly Polarized Light Detection. *Advanced Materials* **2022**, *34* (51), 2204119.

(42) Pan, R.; Tang, X.; Kan, L.; Li, Y.; Yu, H.; Wang, K. Spin-Photogalvanic Effect in Chiral Lead Halide Perovskites. *Nanoscale* **2023**, *15* (7), 3300–3308.

(43) Kauranen, M.; Verbiest, T.; Maki, J. J.; Persoons, A. Second-harmonic Generation from Chiral Surfaces. *The Journal of Chemical Physics* **1994**, *101* (9), 8193–8199.

(44) Maki, J. J.; Verbiest, T.; Kauranen, M.; Elshocht, S. V.; Persoons, A. Comparison of Linearly and Circularly Polarized Probes of Second-order Optical Activity of Chiral Surfaces. *The Journal of Chemical Physics* **1996**, *105* (2), 767–772.

(45) Maki, J. J.; Kauranen, M.; Verbiest, T.; Persoons, A. Uniqueness of Wave-Plate Measurements in Determining the Tensor Components of Second-Order Surface Nonlinearities. *Phys. Rev. B* **1997**, *55* (8), 5021–5026.

(46) Van Elshocht, S.; Verbiest, T.; Kauranen, M.; Persoons, A.; Langeveld-Voss, B. M. W.; Meijer, E. W. Direct Evidence of the Failure of Electric-Dipole Approximation in Second-Harmonic Generation from a Chiral Polymer Film. *The Journal of Chemical Physics* **1997**, *107* (19), 8201–8203.

(47) Sioncke, S.; Verbiest, T.; Persoons, A. Second-Order Nonlinear Optical Properties of Chiral Materials. *Materials Science and Engineering: R: Reports* **2003**, *42* (5–6), 115–155.

(48) Araoka, F.; Ha, N. Y.; Kinoshita, Y.; Park, B.; Wu, J. W.; Takezoe, H. Twist-Grain-Boundary Structure in the B 4 Phase of a Bent-Core Molecular System Identified by Second Harmonic Generation Circular Dichroism Measurement. *Phys. Rev. Lett.* **2005**, *94* (13), 137801.

(49) Yuan, C.; Li, X.; Semin, S.; Feng, Y.; Rasing, T.; Xu, J. Chiral Lead Halide Perovskite Nanowires for Second-Order Nonlinear Optics. *Nano Lett.* **2018**, *18* (9), 5411–5417.

(50) Guo, Z.; Li, J.; Wang, C.; Liu, R.; Liang, J.; Gao, Y.; Cheng, J.; Zhang, W.; Zhu, X.; Pan, R.; He, T. Giant Optical Activity and Second Harmonic Generation in 2D Hybrid Copper Halides. *Angew. Chem. Int. Ed.* **2021**, *60* (15), 8441–8445.

(51) Ristow, F.; Liang, K.; Pittrich, J.; Scheffel, J.; Fehn, N.; Kienberger, R.; Heiz, U.; Kartouzian, A.; Iglev, H. Large-Area SHG-CD Probe Intrinsic Chirality in Polycrystalline Films. *J. Mater. Chem. C* **2022**, *10* (35), 12715–12723.

(52) Spreyer, F.; Mun, J.; Kim, H.; Kim, R. M.; Nam, K. T.; Rho, J.; Zentgraf, T. Second Harmonic Optical Circular Dichroism of Plasmonic Chiral Helicoid-III Nanoparticles. *ACS Photonics* **2022**, *9* (3), 784–792.

(53) Guo, Z.; Li, J.; Liang, J.; Wang, C.; Zhu, X.; He, T. Regulating Optical Activity and

Anisotropic Second-Harmonic Generation in Zero-Dimensional Hybrid Copper Halides. *Nano Lett.* **2022**, *22* (2), 846–852.

(54) Fu, X.; Zeng, Z.; Jiao, S.; Wang, X.; Wang, J.; Jiang, Y.; Zheng, W.; Zhang, D.; Tian, Z.; Li, Q.; Pan, A. Highly Anisotropic Second-Order Nonlinear Optical Effects in the Chiral Lead-Free Perovskite Spiral Microplates. *Nano Lett.* **2023**, *23* (2), 606–613.

(55) Wang, H.; Li, J.; Lu, H.; Gull, S.; Shao, T.; Zhang, Y.; He, T.; Chen, Y.; He, T.; Long, G. Chiral Hybrid Germanium (II) Halide with Strong Nonlinear Chiroptical Properties. *Angew Chem Int Ed* **2023**, *62* (41), e202309600.

(56) Schellman, J. A. Circular Dichroism and Optical Rotation. *Chem. Rev.* **1975**, *75* (3), 323–331.

(57) Kauranen, M.; Van Elshocht, S.; Verbiest, T.; Persoons, A. Tensor Analysis of the Second-Order Nonlinear Optical Susceptibility of Chiral Anisotropic Thin Films. *The Journal of Chemical Physics* **2000**, *112* (3), 1497–1502.

(58) Sioncke, S.; Van Elshocht, S.; Verbiest, T.; Persoons, A.; Kauranen, M.; Phillips, K. E. S.; Katz, T. J. Optical Activity Effects in Second Harmonic Generation from Anisotropic Chiral Thin Films. *The Journal of Chemical Physics* **2000**, *113* (17), 7578–7581.

(59) Han, J.; Guo, S.; Lu, H.; Liu, S.; Zhao, Q.; Huang, W. Recent Progress on Circularly Polarized Luminescent Materials for Organic Optoelectronic Devices. *Advanced Optical Materials* **2018**, *6* (17), 1800538.

(60) Tanaka, H.; Inoue, Y.; Mori, T. Circularly Polarized Luminescence and Circular Dichroisms in Small Organic Molecules: Correlation between Excitation and Emission Dissymmetry Factors. *ChemPhotoChem* **2018**, *2* (5), 386–402.

(61) Sang, Y.; Han, J.; Zhao, T.; Duan, P.; Liu, M. Circularly Polarized Luminescence in Nanoassemblies: Generation, Amplification, and Application. *Adv. Mater.* **2020**, *32* (41), 1900110.

**Supporting Information**

**Magneto-chiral nonlinear optical effect with large anisotropic response in two-dimensional halide perovskite**

Daichi Okada[1]*, Fumito Araoka[1]*

[1] *RIKEN Center for Emergent Matter Science (CEMS), 2-1 Hirosawa, Wako, Saitama 351-0198, Japan*

* To whom correspondence should be addressed.
E-mail: daichi.okada@riken.jp, fumito.araoka@riken.jp

## 1. Materials and measurement

Unless otherwise noted, all reagents and solvents were used as received. (R)- and (S)- 4-Chloro-$\alpha$-methylbenzylamine (R-/S-MBACl) is purchased from Thermo scientific (Alfa Aesar). 57% w/w HI aqueous and 50% $H_3PO_2$ aqueous solution, PMMA are purchased from TCI. PbO and anhydrous DMF are purchased from Kanto Chemical Industry.

Photo-absorption were recorded on a JASCO V-770 spectrophotometer. CD spectrum was measured by JASCO-J1500. X-ray diffraction (XRD) patterns were recorded at 25 °C on RIGAKU model Miniflex600 diffractometers, respectively, with a Cu$K$ radiation source (40 kV and 15 mA). SHG is measured by homemade setup.

## 2. Synthesis of bulk single crystal and preparation of thin film

200 mg PbO powder and 200uL R- or S-MBACl were dissolved in 8 mL HI solution with 1mL $H_3PO_2$ solution as stabilizer at 140°C until the transparent yellow solution is formed. Then the solution is slowly cooled to room temperature (25°C) for 48 hours. The yellow crystals are formed. The obtained crystals are filtrated and washed by toluene, then it was dried in vacuum for more than 24hours.

For preparation of thin film, the synthesized (R- or S-MBACl)$_2$PbI$_4$ crystals were dissolved in DMF at 200mg/ml concentration, the thin film was obtained by spin-coating 5000rpm for 60s and subsequent annealing at 100°C for 10min on quartz substrate. To prevent exposure to the atmosphere, a 30 nm PMMA film was spin-coated on top of the chiral OIHPs film. Thin film preparation is conducted inside a $N_2$ filled glovebox.

## 3, Calculation of $g_{CD}$ value

The recorded CD spectrum transforms into a $g_{CD}$ spectrum using the following equation.

$$g_{CD} = \frac{CD(mdeg)}{32980 \times Absorbance} \quad (1)$$

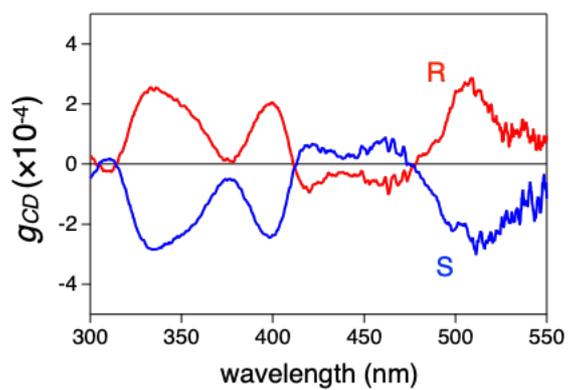

**Figure S1,** (a) $g_{CD}$ spectrum of (R-/S-MBACl)$_2$PbI$_4$ thin film.

## 4, SHG-CD measurement

A Ti:Spphire laser system (Coherent:chameleon) with a pulse width less than 100fs and a repetition rate of 80MHz was used to stimulate the thin film samples. Laser wavelength can be tuned within the range of 1050 to 690nm. Laser source is fixed *p*-polarization by polarizer and sample is irradiated at 45°. In order to generate or detect CP light, achromatic quarter waveplate is placed between sample and polarizer before sample stimulation for SHG-CD measurement or after sample stimulation for CP-SHG measurement. The SHG is collected by a photon counter through a spectrometer (Fig.S2).

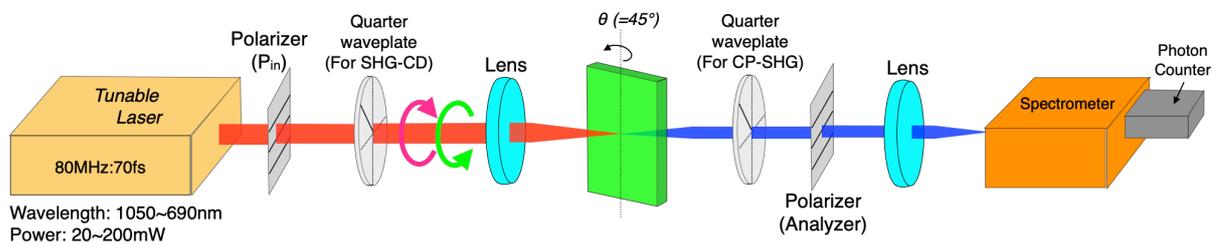

**Figure S2,** Schematic representation of optical se-up for SHG-CD and CP-SHG measurement

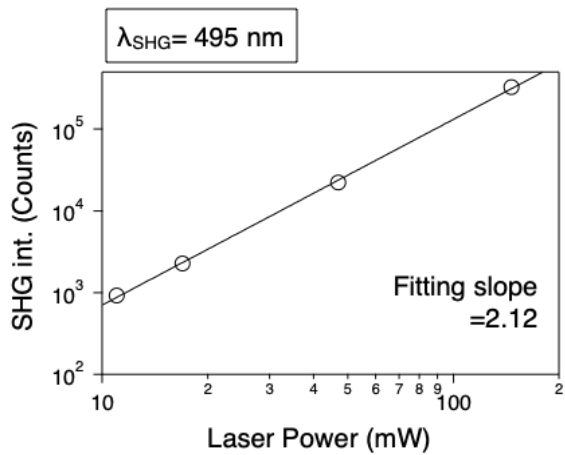

**Figure S3,** Stimulation laser power dependency of SHG intensity at 495nm upon the stimulation of 990nm laser source.

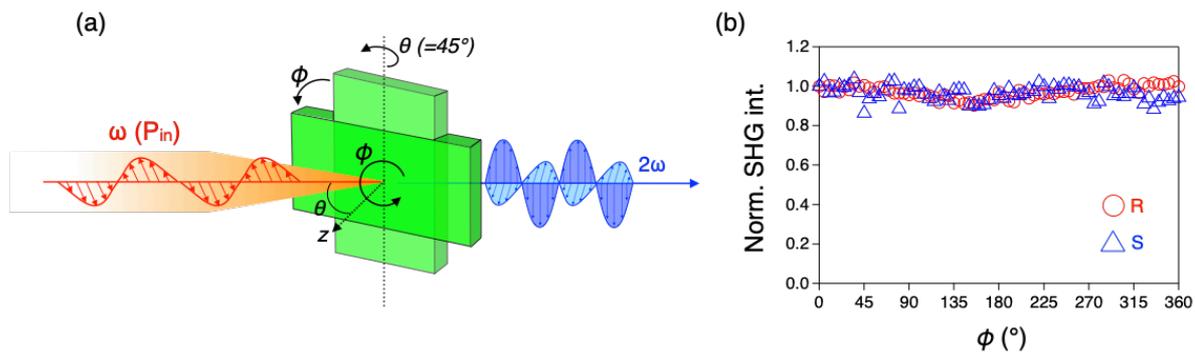

**Figure S4**, (a) Schematic representation of in-plane anisotropy measurement. While rotating the sample in-plane, SHG is recorded. (b) Normalized SHG intensity of R-/S- samples as function of in-plane sample rotation angle upon the stimulation of 990nm laser source.

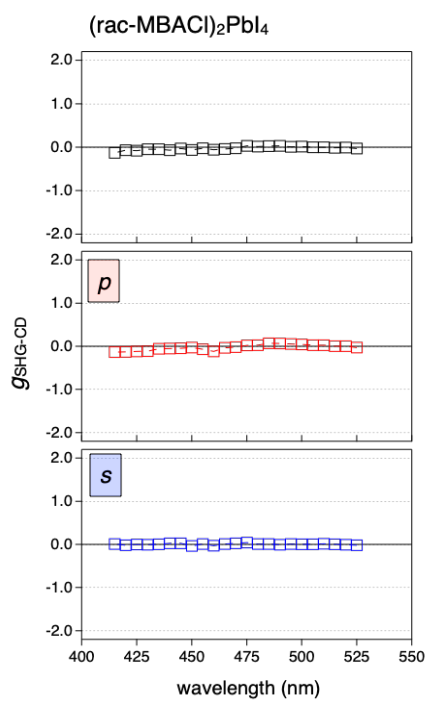

**Figure S5**, $g_{SHG\text{-}CD}$ spectrum of (rac-MBACl)$_2$PbI$_4$ thin film obtained under conditions without a polarizer (Top) and with polarizer for *p*- (Middle) and *s*-polarization (Bottom), respectively.

# 5, Estimation of magnetic dipole (MD) contribution to NLO

MD interaction in NLO process

In nonlinear optical (NLO) processes, the MD interaction can occur in the annihilation of a photon at the fundamental frequency ω or in the creation of a photon with SH frequency 2ω (Fig.S6). In order to include the MD interaction for description of NLO response, we have to consider the MD interaction in nonlinear polarization (2) and nonlinear magnetization (3), and they can be expressed as follows:

$$P_i(2\omega) = \chi_{ijk}^{eee} E_j(\omega) E_k(\omega) + \chi_{ijk}^{eem} E_j(\omega) B_k(\omega) \qquad (2)$$
$$M_i(2\omega) = \chi_{ijk}^{mee} E_j(\omega) E_k(\omega) \qquad (3)$$

where E(ω) and B(ω) are the electric field and magnetic induction fields, respectively. The superscript in the susceptibility component associates the respective subscripts with electric dipole (*e*) and magnetic dipole (*m*) interactions. Both nonlinear polarization and magnetization act as origin of SHG [S1-S5].

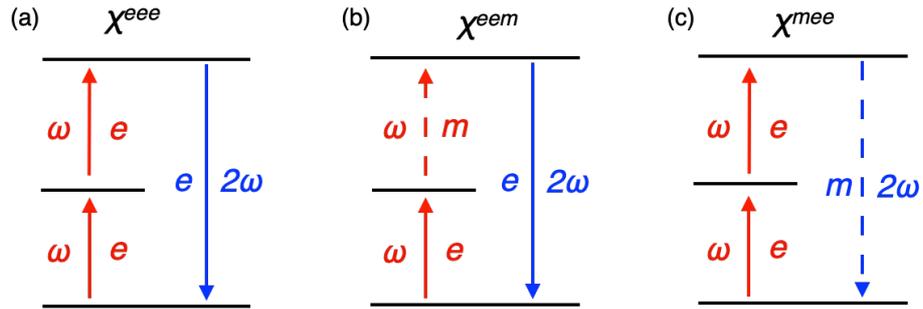

**Figure S6,** (a) ED induction process via the coupling of two EDs (Eq.2, 1st term) (b) ED induction process via the coupling of ED and MD, respectively (Eq.2, 2nd term) (c) MD induction process via the coupling of two EDs (Eq.3)

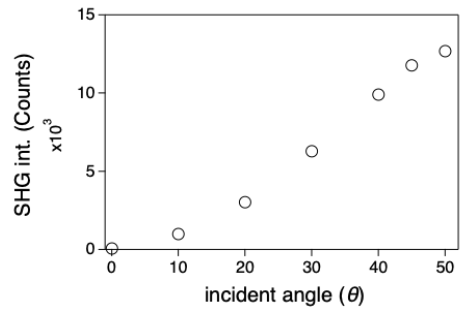

**Figure S7**, Incident laser angle dependency of SHG intensity at $\lambda_{SHG}$=495nm

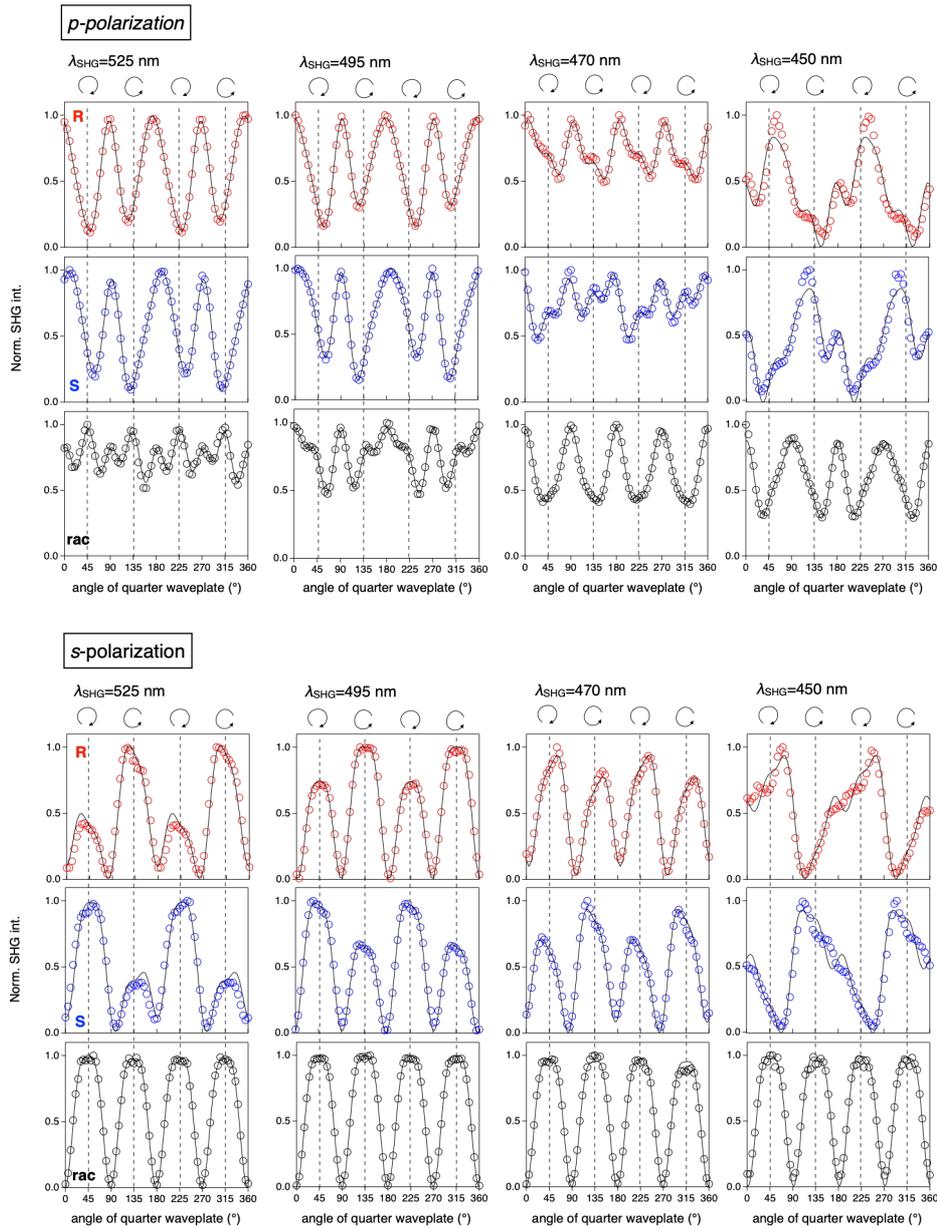

**Figure S8**, Experimental data (dot) and fitting curve (black line) of SHG intensity as function of rotation angle of QWP at several SHG wavelength.

| SHG(nm) | $l_p$(achiral) | $m_p$(achiral) | $n_p$ (chiral) | $l_s$ (chiral) | $m_s$ (chiral) | $n_s$ (achiral) |
|---|---|---|---|---|---|---|
| 525 nm | 1 | 0.03+i0.38 (R)<br>-0.12+i0.52 (S)<br>-1.16-i0.13 (rac) | -(0.29+i0.06)(R)<br>0.37+i0.07 (S)<br>0.10+i0.02(rac) | 0.23+i0.14 (R)<br>-(0.23+i0.14) (S)<br>0.04 (rac) | -(0.18+i0.11)(R)<br>0.25+i0.13 (S)<br>-0.04 (rac) | 1 |
| 495 nm | 1 | -0.28+i0.49(R)<br>-0.44+i0.63 (S)<br>-0.81+i0.41 (rac) | -(0.03+i0.11)(R)<br>0.14+i0.10 (S)<br>0.06-i0.02(rac) | -(0.06-i0.07) (R)<br>0.03-i0.07 (S)<br>0.02 (rac) | 0.06-i0.01 (R)<br>0.02+i0.03 (S)<br>-0.02 (rac) | 1 |
| 470 nm | 1 | -0.64+i0.03(R)<br>-0.77-i0.06 (S)<br>-0.33-i0.10 (rac) | 0.40+i0.04 (R)<br>-(0.36+i0.11) (S)<br>-0.02 (rac) | -(0.22+i0.06) (R)<br>0.21+i0.07 (S)<br>0.03 (rac) | 0.11+i0.03 (R)<br>-(0.07+i0.01) (S)<br>-0.04+i0.01(rac) | 1 |
| 450 nm | 1 | -1.02-i0.84(R)<br>-1.02-i0.91 (S)<br>-0.45-i0.45 (rac) | -0.78+i0.36 (R)<br>-(0.70+i0.32) (S)<br>0.09+i0.02 (rac) | -(0.62+i0.40) (R)<br>0.57+i0.37 (S)<br>-0.06 (rac) | 0.44+i0.16 (R)<br>-(0.36+i0.19) (S)<br>0.07 (rac) | 1 |

**Table S1**, Relative fitting indexes of Figure S7.

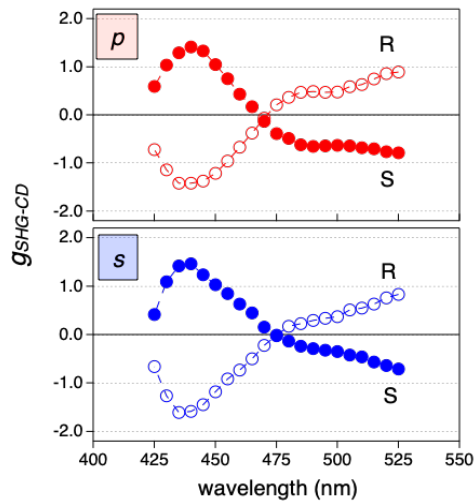

**Figure S9**, The spectrum of dissymmetric factor $g_{SHG\text{-}CD}$ for R/S -samples in *p*- (Top) or *s*-polarized (Bottom) conditions

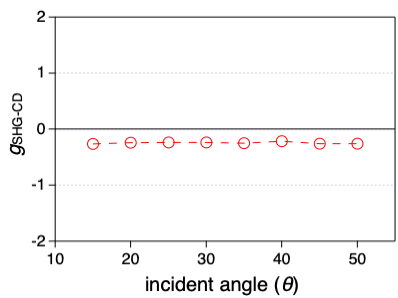

**Figure S10**, Incident laser angle dependency of $g_{SHG\text{-}CD}$ at $\lambda_{SHG}$=495nm

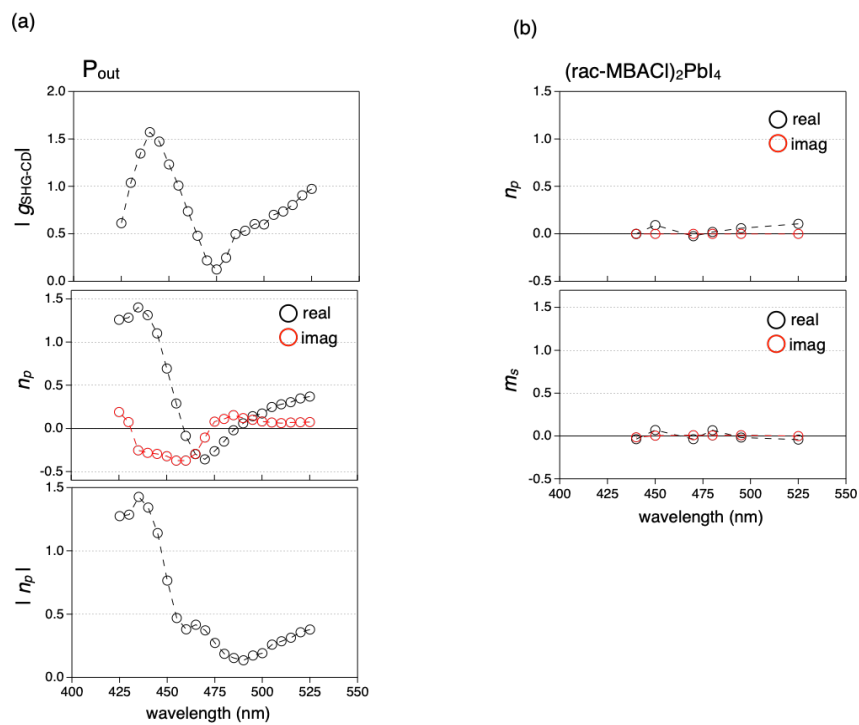

**Figure S11**, (a) Absolute value of $g_{SHG\text{-}CD}$ for (S-MBACl)$_2$PbI$_4$ thin film in *p*-polarized conditions (Top) and spectrum of fitting index $n_p$ with real and imaginary component (Middle) and their absolute value (Bottom). (b) The spectrum of chiral index $n_p$ and $m_s$ for (rac-MBACl)$_2$PbI$_4$ thin film with real and imaginary part.

## 6. Supporting References

(S1) Kauranen, M.; Verbiest, T.; Maki, J. J.; Persoons, A. Second-harmonic Generation from Chiral Surfaces. *The Journal of Chemical Physics* **1994**, *101* (9), 8193–8199.

(S2) Maki, J. J.; Verbiest, T.; Kauranen, M.; Elshocht, S. V.; Persoons, A. Comparison of Linearly and Circularly Polarized Probes of Second-order Optical Activity of Chiral Surfaces. *The Journal of Chemical Physics* **1996**, *105* (2), 767–772.

(S3) Maki, J. J.; Kauranen, M.; Verbiest, T.; Persoons, A. Uniqueness of Wave-Plate Measurements in Determining the Tensor Components of Second-Order Surface Nonlinearities. *Phys. Rev. B* **1997**, *55* (8), 5021–5026.

(S4) Van Elshocht, S.; Verbiest, T.; Kauranen, M.; Persoons, A.; Langeveld-Voss, B. M. W.; Meijer, E. W. Direct Evidence of the Failure of Electric-Dipole Approximation in Second-Harmonic Generation from a Chiral Polymer Film. *The Journal of Chemical Physics* **1997**, *107* (19), 8201–8203.

(S5) Sioncke, S.; Verbiest, T.; Persoons, A. Second-Order Nonlinear Optical Properties of Chiral Materials. *Materials Science and Engineering: R: Reports* **2003**, *42* (5–6), 115–155.